\documentclass[a4paper,10pt,twoside]{cpc-hepnp}

\usepackage{multicol}
\usepackage{graphicx}
\usepackage{booktabs}
\usepackage{amssymb,bm,mathrsfs,bbm,amscd}
\usepackage[tbtags]{amsmath}
\usepackage{lastpage}
\usepackage{CJK}

\begin{document}
%\begin{CJK*}{GBK}{song}

\fancyhead[c]{\small Chinese Physics C~~~Vol. **, No. * (****)
******} \fancyfoot[C]{\small ******-\thepage}

\footnotetext[0]{Received ** ***** ****}

\title{Influence of binding energies of electrons on nuclear mass predictions\thanks{Supported by National Natural Science Foundation of China (11205004) }}

\author{%
      Jing Tang$^{1}$
\quad Zhong-Ming Niu$^{1;1)}$\email{zmniu@ahu.edu.cn}%
\quad Jian-You Guo$^{1}$
}
\maketitle

\address{%
$^1$ School of Physics and Material Science, Anhui University, Hefei 230601, China\\
}

\begin{abstract}
Nuclear mass contains a wealth of nuclear structure information, and has been widely employed to extract the nuclear effective interactions. The known nuclear mass is usually extracted from the experimental atomic mass by subtracting the masses of electrons and adding the binding energy of electrons in the atom. However, the binding energies of electrons are sometimes neglected in extracting the known nuclear masses. The influence of binding energies of electrons on nuclear mass predictions are carefully investigated in this work. If the binding energies of electrons are directly subtracted from the theoretical mass predictions, the rms deviations of nuclear mass predictions with respect to the known data are increased by about $200$ keV for nuclei with $Z, N\geqslant 8$. Furthermore, by using the Coulomb energies between protons to absorb the binding energies of electrons, their influence on the rms deviations is significantly reduced to only about $10$ keV for nuclei with $Z, N\geqslant 8$. However, the binding energies of electrons are still important for the heavy nuclei, about $150$ keV for nuclei around $Z=100$ and up to about $500$ keV for nuclei around $Z=120$. Therefore, it is necessary to consider the binding energies of electrons to reliably predict the masses of heavy nuclei at an accuracy of hundreds of keV.
\end{abstract}

\begin{keyword}
Nuclear masses, binding energies of electrons, Coulomb energies
\end{keyword}

\begin{pacs}
21.10.Dr, 21.60.-n, 21.10.Sf\\
%21.10.Dr Binding energies and masses
%21.60.-n Nuclear structure models and methods
%21.10.Sf Coulomb energies, analogue states
\end{pacs}

\footnotetext[0]{\hspace*{-3mm}\raisebox{0.3ex}{$\scriptstyle\copyright$}2013
Chinese Physical Society and the Institute of High Energy Physics of the Chinese Academy of Sciences and the Institute of Modern Physics of the Chinese Academy of Sciences and IOP Publishing Ltd}%

\begin{multicols}{2}

\section{Introduction}

Nuclear mass is a basic quantity in nuclear physics, and it plays an important role not only in nuclear physics, but also in astrophysics~\cite{Lunney2003RMP, Burbidge1957RMP}. In nuclear physics, it has been widely used to extract the nuclear effective interactions of the mean-field model since it contains a wealth of nuclear structure information, such as the Skyrme effective interactions~\cite{Brown1998PRC, Goriely2009PRLa, Maza2012PRC} and those for the relativistic mean-field (RMF) model~\cite{Meng2006PPNP, Zhao2010PRC, Lalazissis2005PRC, Long2006PLB}. In astrophysics, nuclear mass determines the path of the rapid neuron-capture process ($r$ process), so it is crucial to understand the origin of heavy elements in our universe~\cite{Burbidge1957RMP, Arnould2007PRp, Sun2008PRC, Niu2009PRC, Li2012APS, Xu2013PRC, Sun2015FP}.

The measurement of nuclear masses has achieved great progress in recent years with the development of radioactive ion beam facilities. The latest atomic mass evaluation (AME) was published in 2012 (AME12)~\cite{Wang2012CPC}, and about $200$ new nuclear masses and many more precise nuclear mass data were reported compared with the last version of the AME published in 2003 (AME03)~\cite{Audi2003NPA}. However, for nuclei far from the stability line, the mass measurements of these nuclei are still a great challenge for experimental work. Therefore, theoretical predictions of nuclear masses are inevitable.

During the last few decades, many nuclear mass models have been developed. There are mainly two kinds of nuclear mass models: macro-microscopic and microscopic mass models. The root-mean-square (rms) deviations between the mass predictions of macro-microscopic models and experimental masses in AME12 are generally less than $1$ MeV, such as the finite-range droplet model (FRDM)~\cite{Moller1995ADNDT, Moller2012PRL} and Weizs\"{a}cker-Skyrme (WS)~\cite{Wang2010PRCa, Wang2010PRCb, Liu2011PRC, Wang2014PLB} model. The rms deviation of the WS4 model from experimental masses in AME12 is $298$ keV  for nuclei with $Z, N\geqslant 8$, crossing the $0.3$ MeV accuracy threshold for the first time within the mean-field framework~\cite{Wang2014PLB}. On the other hand, great progress has been made with microscopic mass models as well. A series of Skyrme Hartree-Fock-Bogliubov (HFB) mass models have been developed and their accuracies are comparable with those macro-microscopic models~\cite{Goriely2014PRC}. The model standard deviation of the HFB-27 mass model has been reduced to $0.5$ MeV~\cite{Goriely2013PRC}. In the relativistic framework, the first microscopic mass model was developed in 2005~\cite{Geng2005PTP} with the TMA effective interaction~\cite{Sugahara1994NPA}, and its rms deviation to the known masses in AME12 is about $2.2$ MeV for nuclei with $Z, N\geqslant 8$~\cite{Niu2013PRCb, Zheng2014PRC}. By fitting to the properties of $60$ spherical nuclei, the effective interaction PC-PK1~\cite{Zhao2010PRC} remarkably improves the mass predictions comparing with the TMA effective interaction~\cite{Hua2012SCPMA, Zhao2012PRC, Meng2013FP, Zhang2014FP}, and also successfully describes many other nuclear properties, such as nuclear $\beta$ decay~\cite{Niu2013PRCR}, low-lying excited states~\cite{Li2012PLB, Yao2013PLB, Li2013PLB, Yao2013PRC, Wu2014PRC}, pairing transition at finite temperature~\cite{NiuYF2013PRC}, exotic shape~\cite{ZhaoJ2012PRC}, and the fission barrier~\cite{Lu2014PRC, ZhaoJ2015PRC}.

To determine the effective interactions of nuclear mass models, nuclear masses are the most important data to be fitted. The experimental nuclear mass $m(Z,N)$ is usually determined from the known atomic mass $M(Z,N)$ in AME, which is calculated with
\begin{eqnarray}\label{Eq:m2M}
  m(Z,N) = M(Z,N) - Z\times m_e + B_e(Z),
\end{eqnarray}
where $m_e$ and $B_e(Z)$ are the electron mass and binging energy of electrons in the atom, respectively. The effects of electron screening on the calculated nuclear $\alpha$-decay half-lives~\cite{Dong2010PRC} have been investigated~\cite{Wan2015PRC, Dzyublik2014PRC, Karpeshin2013PRC}. In this work, we will focus on the influence of binding energies of electrons on nuclear mass predictions. The binding energy of electrons in the atom is very sensitive to the proton number of the nucleus. Its contribution in heavy nuclei is remarkable, e.g. this quantity reaches about $760$ keV for uranium. This value can be neglected when the accuracy of the nuclear mass model is about several MeV. In recent days, the accuracy of the nuclear mass model has been remarkably improved and the rms deviation is about $500$ keV. Therefore, the binging energy of electrons is non-negligible in nuclear mass predictions, especially for the heavy nuclei. However, the binding energy of electrons in Eq. (\ref{Eq:m2M}) is still neglected in some nuclear models, such as the RMF~\cite{Geng2005PTP}, Duflo-Zuker (DZ)~\cite{Duflo1995PRC}, and WS~\cite{Wang2010PRCa, Wang2010PRCb, Liu2011PRC, Wang2014PLB} mass models. Therefore, it is interesting to investigate the influence of binding energies of electrons on the nuclear mass predictions. In this work, the influence of the binding energies of electrons is first evaluated by directly considering them in the theoretical mass predictions. Furthermore, we use the Coulomb energies between protons to absorb the binding energies of electrons. In this case, the influence of the binding energy of electrons is reduced, and we discuss this in detail.

\section{Numerical details}

To evaluate the influence of the binding energy of electrons, we will use the rms ($\sigma_{\rm rms} $) and mean ($\varepsilon $) deviations, which are defined to be
\begin{eqnarray}
 \sigma_{rms} &=& \sqrt{\frac{1}{n}\sum\limits_{i = 1}^n (M_i^{\rm th}  - M_i^{\rm exp})^2},\\
 \varepsilon &=& \frac{1}{n}\sum\limits_{i = 1}^n (M_i^{\rm th}  - M_i^{\rm exp}),
\end{eqnarray}
where $M_i^{\rm th} $ and $M_i^{\rm exp} $ are the theoretical and experimental atomic masses, and $n$ is the number of atoms contained in a given set.

The experimental atomic masses are taken from AME12~\cite{Wang2012CPC} and the formula of the binding energy of electrons,
\begin{eqnarray}\label{Eq:Be}
 B_e(Z) = 14.4381 \times {Z^{2.39}} + 1.55468 \times {10^{ - 6}} \times {Z^{5.35}}~{\rm{eV}},
\end{eqnarray}
is adopted~\cite{Lunney2003RMP, Wang2012CPC}. The theoretical masses are taken from the nuclear models. In this work, the RMF~\cite{Geng2005PTP}, DZ28~\cite{Duflo1995PRC}, and WS4~\cite{Wang2014PLB} models are employed to investigate the influence of binding energy of electrons.

\section{Results and discussion}

\begin{center}
\tabcaption{ \label{Tab:rmsepsi} The rms and mean deviations of mass predictions with respect to the known masses in AME12 for various models. The second and third (fourth and fifth) columns represent the results for the group of nuclei with $Z, N\geqslant 8$ ($Z\geqslant 60$).}
\footnotesize
\begin{tabular*}{73mm}{l@{\extracolsep{\fill}}cccc}
\toprule
        &\multicolumn{2}{c}{$Z, N\geqslant 8$}  &\multicolumn{2}{c}{$Z\geqslant 60$}\\
%\hline
Model   &$\sigma_{\rm rms}$   &$\varepsilon$              &$\sigma_{\rm rms}$   &$\varepsilon$\\
\hline
RMF           &2.217    &-0.788     &2.583    &-1.560 \\
RMF*          &2.441    &-1.090     &2.961    &-2.085 \\
RMF\#         &2.220    &-0.759     &2.592    &-1.563 \\
DZ28          &0.394    &-0.032     &0.366    &-0.035 \\
DZ28*         &0.583    &-0.342     &0.704    &-0.572 \\
DZ28\#        &0.398    &-0.006     &0.375    &-0.043 \\
WS4           &0.298    &-0.003     &0.242    &-0.004 \\
WS4*          &0.501    &-0.312     &0.618    &-0.540 \\
WS4\#         &0.304    &~0.024     &0.248    &-0.011 \\
\bottomrule
\end{tabular*}
\end{center}

For comparison with the experimental atomic mass, the calculated nuclear mass $m^{\rm th}(Z,N)$ is usually transformed to the atomic mass $M^{\rm th}(Z,N)$. If the binding energy of electrons is neglected as in the RMF, DZ, and WS mass models, one then has
\begin{eqnarray}
  M^{\rm th}(Z,N) = m^{\rm th}(Z,N) + Z\times m_e.
\end{eqnarray}
By including the binding energy of electrons, the revised atomic mass $M^*_{\rm th}$ should be
\begin{eqnarray}
M_*^{\rm th}(Z, A)= M^{\rm th}(Z, A) - B_e(Z).
\end{eqnarray}
To distinguish them from the original mass models, the revised mass models are denoted with Model* hereafter. In Tab. (\ref{Tab:rmsepsi}), the rms and mean deviations of the RMF, DZ28, and WS4 models and the corresponding results after considering the binding energies of electrons, i.e. mass predictions of the RMF*, DZ28*, and WS4* models, are given. For the group of nuclei with $Z, N\geqslant 8$, the rms deviations of the RMF*, DZ28*, and WS4* models are $224$ keV, $189$ keV, and $203$ keV larger than those of the RMF, DZ28, and WS4 models, respectively. In addition, the mean deviations of the RMF*, DZ28*, and WS4* models are about $300$ keV smaller than the results without considering $B_e$. This indicates the binding energy of electrons has an important impact on the accuracy of nuclear mass predictions, especially for those models with rms deviations of hundreds of keV. Since the binding energy of electrons monotonically increases with the proton number, its influence on the mass predictions is certainly increased for the heavy nuclei with larger proton numbers. Therefore, the rms and mean deviations for nuclei with $Z\geqslant 60$ are also given in Tab. 1. Clearly, the influence of the binding energy of electrons on the rms and mean deviations is significantly increased, and its influence on rms and mean deviations is about $350$ keV and $500$ keV, respectively.

\begin{center}
\includegraphics[width=7cm]{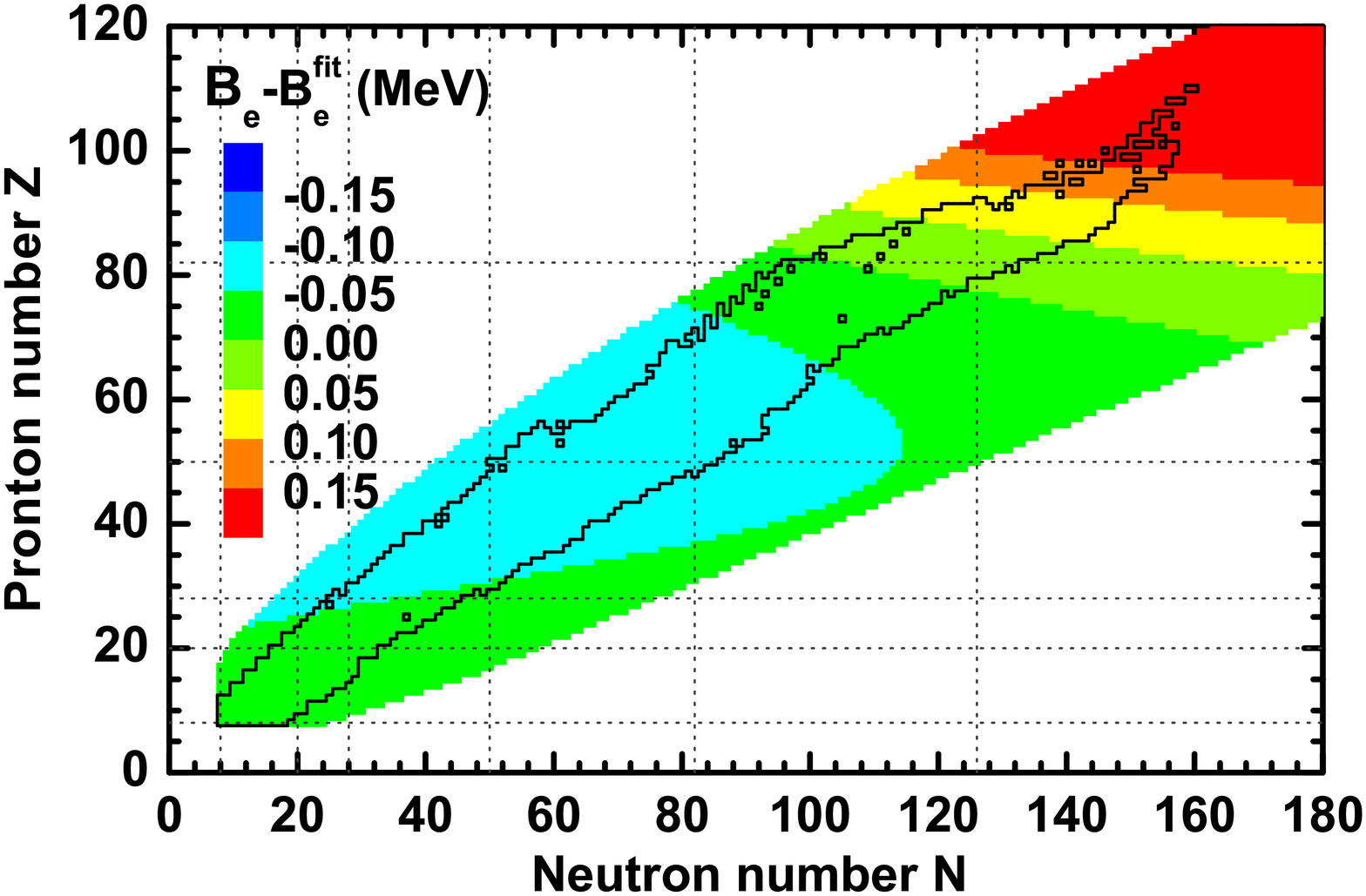}
\figcaption{\label{Fig:BeSubBefit} (Color online) The difference between $B_e$ and $B_e^{\rm fit}$ for the nuclei in the WS4 model. Boundaries of nuclei with known masses in AME12 are shown by the black contours.}
\end{center}

For those models which do not consider the binding energy of electrons, their effective interactions are usually determined by fitting to the data $M^{\rm exp}(Z,N)-Z\times m_e=m^{\rm exp}(Z,N)-B_e$, where the binding energies of electrons are subtracted from known nuclear masses. Therefore, the subtraction of binding energies of electrons may be partially compensated by adjusting their effective interactions. Both the binding energy of electrons $B_e$ and the Coulomb energy $E_C$ between protons originate from the Coulomb interaction between charged particles, so the formulas describing these two kinds of energies are mainly dependent on the proton number $Z$, although $E_C$ has a weak dependence on mass number $A$. Therefore, the binding energy of electrons $B_e$ may be compensated partially by fitting to the data with Coulomb energy $E_C$ between protons. In this work, we take the formula of Coulomb energy in the WS4 model as an example, which is
\begin{eqnarray}\label{Eq:Ec}
E_C(Z, A) = a \times \frac{{{Z^2}}}{{{A^{\frac{1}{3}}}}} \times \left( {1 - 0.76 \times {Z^{ - \frac{2}{3}}}} \right) .
\end{eqnarray}
The parameter $a$ in Eq. (\ref{Eq:Ec}) is determined by fitting Eq. (\ref{Eq:Be}) to $B_e$ for all known nuclei in AME12 with $Z, N\geqslant 8$. Then the revised mass $M_\#^{\rm th}$ is calculated with
\begin{eqnarray}
M_\#^{\rm th}(Z, A)=M^{\rm th}(Z, A)-[B_e(Z)-B_e^{\rm fit}(Z, A)] ,
\end{eqnarray}
where $B_e^{\rm fit} $ is calculated using Eq. (\ref{Eq:Ec}) with the fitted value $a=0.505$ keV. The revised mass models are denoted with Model\# for simplicity. In Fig.~\ref{Fig:BeSubBefit}, we show the differences between $B_e$ and $B_e^{\rm fit}$ for various nuclei in the WS4 model. From Fig.~\ref{Fig:BeSubBefit}, it is found that $B_e^{\rm fit} $ is smaller than $B_e$ when $Z\lesssim 80$, while it becomes greater than $B_e$ when $Z\gtrsim 80$. The differences between $B_e$ and $B_e^{\rm fit} $ are almost within $100$ keV for nuclei with $Z\lesssim 100$. However, the value of $B_e-B_e^{\rm fit}$ increases significantly when $Z\gtrsim 100$, being about $150$ keV for nuclei around $Z=100$ and up to about $500$ keV for nuclei around $Z=120$. This indicates that the formula of Coulomb energy between protons indeed significantly compensates the subtraction of binding energy of electrons for nuclei with $Z\lesssim 100$, while its influence on the mass predictions of heavy nuclei with $Z\gtrsim 100$ is still important. Since there are few known nuclei with $Z\gtrsim 100$ in AME12, the influence of binding energy of electrons on the rms deviations would be small. In Tab.~\ref{Tab:rmsepsi}, the rms and mean deviations of the DZ28\#, RMF\#, and WS4\# models are given as well. Comparing with those values of the RMF*, DZ28*, and WS4* models, the influence of binding energy of electrons is remarkably reduced. The changes of the rms deviations with respect to original values of the DZ28, RMF, and WS4 models are within $10$ keV, even for the group of nuclei with $Z\geqslant 60$.

\end{multicols}
%\ruleup
\begin{center}
\includegraphics[width=15cm]{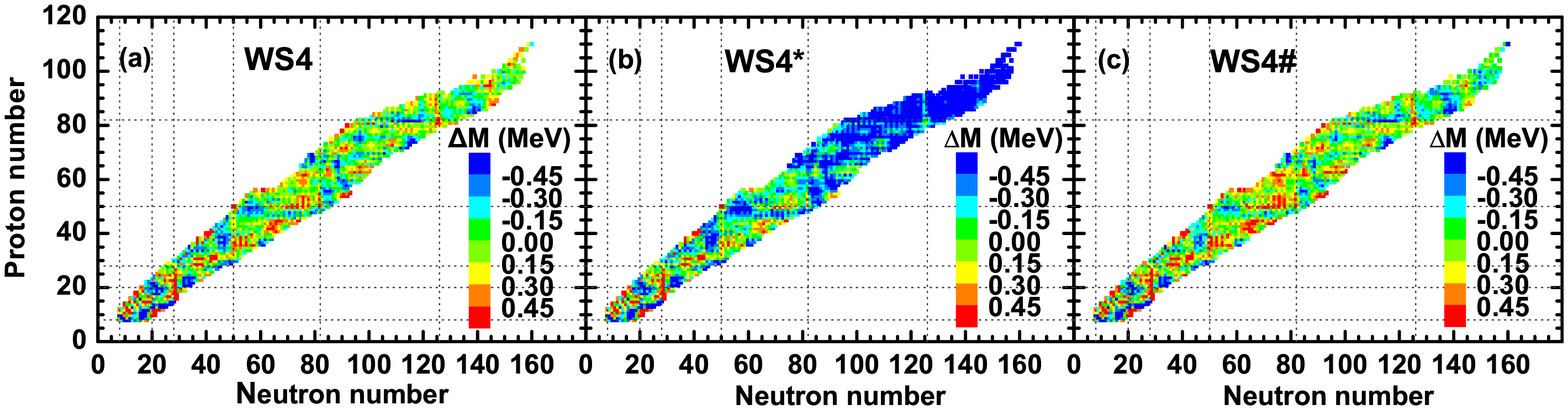}
\figcaption{\label{Fig:DMWS4WS4sWS4p} (Color online) Mass differences between theoretical predictions and the experimental data. The results for the WS4, WS4*, and WS4\# mass models are shown in panels (a), (b), and (c), respectively.}
\end{center}
%\ruledown
\begin{multicols}{2}

To better study the influence of binding energy of electrons, the mass differences between the theoretical predictions and the experimental data are shown on the nuclear chart in Fig.~\ref{Fig:DMWS4WS4sWS4p} for the WS4, WS4*, and WS4\# mass models. It is clear that the mass differences of WS* model are significantly different from those of WS model, especially for nuclei with $Z\gtrsim 60$. This can be understood since the value of $B_e$ for nucleus with $Z\geqslant 60$ is even larger than the rms deviation of WS4 model for nuclei with $Z\geqslant 60$. By using the Coulomb energies between protons to absorb the binding energies of electrons, the mass differences of the WS4\# model are generally similar to those of the WS4 model, although mass differences of the WS4\# model are slightly increased for nuclei around $Z=50$. In addition, the mass differences decrease significantly due to the large value of $B_e-B_e^{\rm fit}$ (see Fig.~\ref{Fig:BeSubBefit}) when $Z\gtrsim 100$. This further indicates the importance of the influence of binding energies of electrons on the mass predictions of heavy nuclei.

\section{Summary}

In this work, the influence of binding energies of electrons on nuclear mass predictions is carefully investigated. It is found that the rms deviations of nuclear mass predictions with respect to the known data are increased by about $200$ keV for nuclei with $Z, N\geqslant 8$ by directly subtracting the binding energies of electrons from the theoretical mass predictions. Since both the Coulomb energies between protons and the binding energies of electrons originate from the Coulomb interactions between charged particles and are determined by the proton number, we further use the Coulomb energies between protons to absorb the binding energies of electrons. It is found that the Coulomb energies between protons can absorb the binding energies of electrons well for known nuclei in AME12, with the differences of rms deviations only being about $10$ keV compared with the original values. However, the binding energies of electrons are still important for the heavy nuclei, about $150$ keV for nuclei around $Z=100$ and up to about $500$ keV for nuclei around $Z=120$. The studies in this work imply that the inclusion of the binding energies of electrons to extract the nuclear effective interactions would help to improve the mass predictions of heavy nuclei with $Z\gtrsim 100$.

\vspace{3mm}
\acknowledgments{We thank Dr. Bao-Hua Sun and Dr. Yi-Fei Niu for stimulating suggestions.}

\end{multicols}

\clearpage

%\end{CJK*}
\end{document}